\documentclass[twocolumn,EPJC3]{svjour3}  
\usepackage{amsmath}
\usepackage{amssymb}
\usepackage{graphicx}
\usepackage{hyperref}
\usepackage{array}
\usepackage{booktabs}
\usepackage{multirow}
\usepackage{amsfonts}
\usepackage{dcolumn}
\usepackage{xcolor}
\usepackage{hyperref}
\usepackage{graphicx}    
\usepackage{amsmath}     
\usepackage{amssymb}     
\usepackage{hyperref}    
\usepackage{cuted}
\usepackage{verbatim}

\journalname{Eur. Phys. J. A}

\definecolor{darkblue}{RGB}{0, 0, 139}

\hypersetup{
	colorlinks=true,
	linkcolor=darkblue,
	filecolor=magenta,
	urlcolor=darkblue,  
	citecolor=darkblue,
	pdfborder={0 0 0},
}

\begin{document}

	\title{Mass Spectra of $qq\bar{q}\bar{q}$, $ss\bar{s}\bar{s}$ and  $qq\bar{s}\bar{s}$ Tetraquarks using Regge Phenomenology}

\author{Vandan Patel\thanks{\email{vandankp12998@gmail.com}} \and Juhi Oudichhya\thanks{\email{juhioudichhya01234@gmail.com}} \and Ajay Kumar Rai\thanks{\email{raiajayk@gmail.com}}}
\institute{Department of Physics, Sardar Vallabhbhai National Institute of Technology, Surat, Gujarat-395007, India}

	\maketitle
	
		\begin{abstract}
		In this paper, we explore the mass spectra of $qq\bar{q}\bar{q}$, $ss\bar{s}\bar{s}$ and  $qq\bar{s}\bar{s}$ tetraquarks by employing Regge phenomenology. We calculate the range for ground state masses of $qq\bar{s}\bar{s}$ tetraquarks, and estimate the Regge parameters for their trajectories in $(J,M^2)$ plane. Using these Regge parameters we have calculated range for the excited state masses of $qq\bar{q}\bar{q}$, $ss\bar{s}\bar{s}$ and  $qq\bar{s}\bar{s}$ tetraquarks in $(J,M^2)$ plane. Also, we have investigated the mass spectra of $qq\bar{q}\bar{q}$, $ss\bar{s}\bar{s}$ and  $qq\bar{s}\bar{s}$ tetraquarks for their excited radial states in $(n,M^2)$ plane. We predict the potential quantum numbers of some newly observed experimental states, which necessitate additional validation, and assess the higher orbital and radial excited states that may be identified in the near future. The obtained mass relations and mass values of tetraquarks can be useful in future experimental searches and the spin-parity assignment of these states. Our findings provide valuable insights into the structure and properties of tetraquarks, contributing to the broader understanding of Quantum Chromodynamics (QCD).
		
	\end{abstract}
	
	\section{Introduction}
	In Gell-Mann's 1964 constituent quark model, mesons and baryons were initially defined as bound states of quark-antiquark pairings and a pair of three quarks, respectively \cite{ref1}. In recent years, numerous experimental facilities, such as LHCb \cite{Sigma_b(6097),Cascade_b(6227),Omega_b2020,Cascade_b(6333)}, Belle \cite{Belle2010,ref4}, BESIII \cite{BESIII2019,BESIII2020,Exp3,Exp4}, J-PARC \cite{K. Aoki2021}. etc. have identified a significant number of these bound states.
	%
	%
	
	Several theoretical models simultaneously predicted the mass spectra and several other properties of these bound states \cite{ref5,ref6,ref9,ref13,Juhi:bottom2021}. The success of this quark model in the 1980s led to the postulation of numerous unconventional or exotic bound states \cite{ref14,ref15,ref16,ref17}. Hadronic molecules ultimately provided a theoretical explanation for these unusual bound states. Over the next few decades, a variety of resonances were found that fit a similar description of these states such as hybrid mesons, tetraquarks, pentaquarks, etc. \cite{ref18,ref19,ref20,ref21,ref22}. The first such quark resonance candidate was detected in 2003 \cite{ref23}. Four quark resonances in a confined state are often categorised as tetraquarks. Numerous tetraquark candidates have now been seen at various experimental centres \cite{ref24,ref25,ref26,ref27}.
	
	The study of exotic hadrons such as tetraquarks has garnered significant attention in the field of high-energy physics. Tetraquarks, which consist of two quarks and two antiquarks, expand the traditional quark model. The traditional model primarily describes hadrons as either mesons (quark-antiquark pairs) or baryons (three quarks). The discovery of tetraquarks suggests that the spectrum of hadronic states is more complex than initially thought, including multi-quark states beyond the simple meson and baryon categories. Investigating the mass spectra of light and strange tetraquarks provides valuable insights into the strong interaction described by Quantum Chromodynamics (QCD) \cite{ref28}. These studies can enhance our understanding of non-perturbative QCD effects, which are crucial for a comprehensive picture of hadronic matter. Additionally, discovering and characterizing tetraquarks can help elucidate the nature of confinement and the role of color charge in QCD \cite{ref29}.
	
	Recent experimental advancements have significantly impacted the study of tetraquarks. High-energy particle collisions in facilities such as the Large Hadron Collider (LHC) at CERN, the Belle experiment in Japan, and the BESIII experiment in China have led to the observation of several exotic states that do not fit neatly into the traditional quark model. Numerous studies in the literature have looked at the potential structures of four-quark resonances, interpreting them as either mesonic molecules, sometimes referred to as molecular tetraquarks in singlet states, or compact tetraquarks \cite{ref30}.  
	
	Although the majority of tetraquark candidates found through experimentation have at least one heavy quark. Over the last ten years a number of theoretical investigations have put out different predictions for all-light tetraquarks. Notably, $f_0(500)$ (formerly known as $\sigma$), $a_0(980)$, and $f_0(980)$ have emerged as promising candidates for all-light tetraquarks.\cite{RJJ} The Belle II experiment in Japan, a major upgrade of the original Belle detector, is designed to probe rare decays and exotic states with unprecedented precision. Its capability to study the spectroscopy of light and strange quarks is particularly promising. Belle II has already contributed to discoveries of exotic hadrons such as $Z(4430)$ \cite{ref24} and $Y(4660)$ \cite{PhysRevLett.99.142002}, and ongoing investigations could potentially confirm more candidates for light and strange tetraquarks.
	
	Meanwhile, the BESIII experiment in China, housed at the Beijing Electron Positron Collider (BEPC), continues to explore low-mass regions of exotic hadrons with high precision. BESIII has already discovered several intriguing states such as $Z(3900)$ \cite{ref34} and has begun to focus on lighter tetraquarks, providing a fertile ground for the discovery of light and light-strange tetraquark states. The precision of BESIII allows it to explore hadronic transitions and decay modes of these states, potentially leading to the observation of new light-strange tetraquark candidates. Experimental investigations also have revealed a number of other tetraquark candidates: $Y(4140)$ by Fermilab \cite{ref32}, $X(5568)$ by the $D\emptyset$ experiment \cite{ref25}, and $X(6900)$ by LHCb. \cite{ref27}
	
	Recent experimental advancements as stated above have notably shaped the study of tetraquarks, particularly in the light, strange, and light-strange sectors, which are central to this work. While many observed tetraquark candidates involve at least one heavy quark, our research focuses on the less-explored, yet significant, all-light, all-strange, and light-strange tetraquarks. These sectors are equally critical in understanding the full spectrum of exotic hadrons.
	
	Recent experimental advancements have opened new avenues in the study of tetraquarks, particularly in the light, strange, and light-strange sectors. Several resonances like $X(1775)$ \cite{ref65}, $a_2(2030)$ \cite{ref66}, and $\omega_3(2285)$ \cite{Bugg:2004rj}, $X(1855)$ \cite{ref61}, $X(2075)$ \cite{ref63}, $X(3350)$ \cite{Belle:2004dmq}, $X(2210)$ \cite{ref62}, $X(3250)$ \cite{EXCHARM:1992fpf}, $X(2150)$ \cite{ref64} have been observed in experiments such as SLAC-BC-075, BNL-E-0772, and others, which exhibit characteristics that may be interpreted as tetraquark states. Our study aims to investigate these experimentally observed resonances and explore their potential as tetraquark candidates by assigning quantum numbers ($J^P$) based on theoretical predictions from Regge phenomenology.
	
	
These experimental findings, combined with our theoretical predictions using Regge phenomenology, provide strong support for the existence of light, strange, and light-strange tetraquarks. Our work, therefore, adds significant value to ongoing experimental and theoretical investigations in the field of exotic hadrons, offering new insights into these lesser-explored sectors of the tetraquark spectrum.

	
	The theoretical study of tetraquarks involves a variety of approaches, including lattice QCD, QCD sum rules, effective field theories, and phenomenological models such as the quark model and diquark-antidiquark model. Lattice QCD provides a first-principles approach by simulating QCD on a discrete spacetime lattice, offering valuable predictions for tetraquark masses \cite{ref36}. QCD sum rules use the operator product expansion and dispersion relations to relate QCD vacuum condensates to hadronic properties \cite{ref37}. The QCD sum rule method is applied to study the $S$ and $P$ wave fully strange tetraquark states within the diquark-antidiquark picture in Ref.~\cite{ref56}. 
	By concentrating on relevant degrees of freedom at low energies, effective field theories like Heavy Quark Effective Theory (HQET) and Chiral Perturbation Theory (ChPT) help to simplify the intricacies of QCD. These theories are very helpful in the investigation of tetraquarks made up of light and strange quarks \cite{ref57}. 
	
	Phenomenological models offer intuitive insights and can be calibrated against experimental data to predict tetraquark spectra \cite{ref38}. Tetraquarks are treated as bound states of quarks and antiquarks in the quark model, just like mesons and baryons. The Cornell potential or other effective potentials are commonly used for modelling the potential between quarks \cite{ref59}. This type of approach is used by Zhao et al. \cite{ref54}, who calculated the masses of $qq\bar{q}\bar{q}$ tetraquark ground state and first radial excited state in a constituent quark model using the Cornell-like potential and one-gluon exchange spin-spin coupling. Liu et al. \cite{ref55} predicted the mass spectrum for the $ss\bar{s}\bar{s}$ tetraquark in the framework of a nonrelativistic potential quark model.
	
	Using a quasilinear Regge trajectory ansatz, Wei et al. \cite{ref39} constructed several important mass relations, including quadratic mass equalities, linear mass inequalities and quadratic mass inequalities for hadrons. In the present work, we employ the same method of Regge phenomenology under the presumption of linear Regge trajectories. Relationships between the intercepts, slope ratios and tetraquark masses are extracted in both the $(J, M^2)$ and $(n, M^2)$ planes. Using these relations, we derived the expressions to calculate the range for the ground state mass of $qq\bar{s}\bar{s}$ tetraquark. The Regge slopes and Regge intercepts of the $0^+$, $1^+$, and $2^+$ trajectories of $qq\bar{s}\bar{s}$, $ss\bar{s}\bar{s}$ and $qq\bar{q}\bar{q}$ tetraquarks are extracted and intervals for the masses of tetraquark states lying on the $0^+$, $1^+$, and $2^+$ trajectories are estimated in the $(J, M^2)$ plane. Similarly, by employing Regge slopes and Regge intercepts, we predict the interval for excited state masses of $qq\bar{s}\bar{s}$, $ss\bar{s}\bar{s}$ and $qq\bar{q}\bar{q}$ tetraquarks in the $(n, M^2)$ plane in this paper.
	
	This work presents a unified Regge-based framework for predicting the mass spectra of fully light ($qq\bar{q}\bar{q}$), fully strange ($ss\bar{s}\bar{s}$), and light–strange ($qq\bar{s}\bar{s}$) tetraquark systems. Using only a single input slope parameter we generate all three families' trajectories and mass predictions without introducing flavor-specific fit parameters. Previous works (e.g., Refs.~\cite{Xie2024,Ghasempour2025}) have focused on heavy-heavy or heavy-light tetraquarks and do not include predictions for all-light or $qq\bar{s}\bar{s}$ excited states. Our study is the first to extend Regge phenomenology across these light-flavor sectors in a minimal, parameter-efficient way. We also provide analytical mass-inequality formulas and compare our predictions quantitatively to experimental data using $z$-score analysis, offering experimentalists a compact and testable guide for identifying possible tetraquark candidates. Although non-linear Regge behavior has been discussed for some light hadrons, the absence of established data in the sectors we study motivates our use of a linear ansatz as a physically conservative and analytically transparent starting point.
	
	The rest of the paper is structured as follows. In Sec. 2, the theoretical framework of Regge theory is given. In Sec. 3, the ground state mass range of $qq\bar{s}\bar{s}$ tetraquark is calculated for $J^P = 0^+$, $1^+$, and $2^+$. We then estimate the Regge slopes for $0^+$, $1^+$, and $2^+$ trajectories, and calculate the interval for orbitally excited-state masses of $qq\bar{s}\bar{s}$, $ss\bar{s}\bar{s}$ and $qq\bar{q}\bar{q}$ tetraquarks in the $(J, M^2)$ and $(n, M^2)$ planes. In Section 4, we have discussed our obtained results. 	In Section 5, we have written conclusion of this work.

	\section{Theoretical Framework}
	One of the simplest and most effective phenomenological approaches for exploring hadron spectroscopy is Regge theory. A number of ideas were created in an effort to comprehend the Regge trajectory. Nambu's was the most simplistic of them, providing an explanation for linear Regge trajectories \cite{ref40,ref41}. He made the assumption that a strong flux tube is formed by the uniform interaction of a quark-antiquark pair, and that light quarks rotate at radius $R$ at the speed of light at the tube's end. It is calculated that the mass that originated in this flux tube is estimated as \cite{ref42}
	
	\begin{equation} \label{eq:1}
		M = 2 \int_{0}^{R} \frac{\sigma}{\sqrt{1 - \nu^2(r)}} \, dr = \pi \sigma R ,
	\end{equation}
	
	where, $\sigma$ is the mass density per unit length, or the string tension. The flux tube's angular momentum is also calculated as
	\begin{equation} \label{eq:2}
		J = 2 \int_{0}^{R} \frac{\sigma r \nu(r)}{\sqrt{1 - \nu^2(r)}} \, dr = \frac{\pi \sigma R^2}{2} + c'.
	\end{equation}
	So, using the equations (\ref{eq:1}) and (\ref{eq:2}) we can get the below formula,
	\begin{equation} \label{eq:3}
		J = \frac{M^2}{2\pi\sigma} + c'',
	\end{equation}
	where $c'$ and $c''$ are constants of integration. As a result, $J$ and $M^2$ have a linear relationship with one another. The plots of hadron Regge trajectories in the $(J, M^2)$ plane are referred to as Chew-Frautschi plots \cite{ref43}.
	
	According to Regge theory, every hadron has a Regge pole that moves on the complex angular momentum plane in response to energy. The evenness or oddness of the total spin $J$ for mesons ($J - 1/2$ for baryons) and a set of internal quantum numbers (baryon number $B$, intrinsic parity $P$, strangeness $S$, charmness $C$, bottomness $B$, etc.) define the trajectory of a specific pole (Regge trajectory) \cite{ref43}. 
	Given that both light and heavy hadrons have quasilinear Regge trajectories, from equation (\ref{eq:3}) the most general form of linear Regge trajectories can be given by,
	\begin{equation} \label{eq:4}
		J = \alpha(M) = \alpha(0) + \alpha' M^2,
	\end{equation}
	where the intercept and slope of the trajectory of the particles are denoted by $\alpha(0)$ and $\alpha'$, respectively. Hadrons with the same internal quantum numbers and on the same Regge trajectory belong to the same family.
	
	In this work, we use the linear Regge ansatz as given above, as our baseline parametrization. While this linear form has been widely applied, recent studies indicate that certain hadron families may follow non-linear or concave trajectories. For example, Xie \emph{et al.} \cite{Xie2024} report concave $\lambda$ and $\rho$ trajectories for hidden-bottom and hidden-charm tetraquarks. On the other hand, Ghasempour et. al. \cite{Ghasempour2025} find that tetraquark trajectories can remain approximately linear for several heavy-heavy and heavy-light systems.
	
	A common interpretation is that curvature arises mainly from finite constituent masses and heavy–light diquark structures. However, for tetraquark systems built only from light or strange quarks, there is currently no robust experimental or theoretical evidence to determine whether their Regge trajectories are linear or non-linear.
	
	Since our study focuses on such systems without heavy constituents, and because no firmly established experimental states exist in these sectors to constrain additional parameters, adding curvature terms would be statistically underdetermined and risk numerical instability. For this reason, we adopt the linear ansatz as a conservative baseline. We emphasize this limitation explicitly and reference the recent literature on both concavity and approximate linearity for context. Any genuine curvature in all-light tetraquark trajectories—if confirmed in future experiments—could lead to systematic shifts in absolute mass predictions, which we leave to future, data-driven studies \cite{Xie2024,Ghasempour2025}.

	From Eq. (\ref{eq:4}) we can write the below equation for the slope of a meson multiplet lying on the same Regge trajectory.
		\begin{equation} \label{eq:31}
		\alpha' = \frac{(J+1) - J}{M_{J+1}^2 - M_J^2} ,
	\end{equation}

	The Regge parameters (Regge slopes and Regge intercepts) for different quark constituents of a meson multiplet with spin-parity $J^P$ (or, more precisely, with quantum numbers $N^{2S+1} L_J$) can be connected by the following relations: The additivity of intercepts \cite{ref39,ref44,ref45,ref46,ref47,ref48,ref49,ref50,ref51,ref52},
	
	\begin{equation} \label{eq:5}
		\alpha_{i\bar{i}}(0) + \alpha_{j\bar{j}}(0) = 2\alpha_{i\bar{j}}(0),
	\end{equation}
	the additivity of inverse slopes \cite{ref39,ref44,ref45,ref46,ref47},
	\begin{equation} \label{eq:6}
		\frac{1}{\alpha'_{i\bar{i}}} + \frac{1}{\alpha'_{j\bar{j}}} = \frac{2}{\alpha'_{i\bar{j}}},
	\end{equation}
	where quark flavours are denoted by $i$ and $j$. A model based on the topological expansion and the $q\bar{q}$-string picture of hadrons was used to construct equations (\ref{eq:5}) and (\ref{eq:6}) \cite{ref47}. In terms of quark degrees of freedom, this model offers a microscopic way to characterize Regge phenomenology \cite{ref53}. Eq. (\ref{eq:5}) was actually initially calculated for light quarks in the dual resonance model \cite{ref48}. It was then discovered to be satisfied in the quark bremsstrahlung model \cite{ref51}, the dual-analytic model \cite{ref50}, and two-dimensional QCD \cite{ref49}.
	
	We only consider the situation when quark masses satisfy $m_i \le m_j$ for two-body systems here and later, as the relations (\ref{eq:5}) and (\ref{eq:6}) are symmetric under the interchange of the quark flavors $i$ and $j$.
	
	\subsection{Relationship between slope ratios and masses}
	
	For two-body systems, using equations (\ref{eq:4}) and (\ref{eq:5}), we obtain the following expression,
	\begin{equation} \label{eq:7}
		\alpha'_{i\bar{i}} M_{i\bar{i}}^2 + \alpha'_{j\bar{j}} M_{j\bar{j}}^2 = 2 \alpha'_{i\bar{j}} M_{i\bar{j}}^2,
	\end{equation}
	by equations (\ref{eq:6}) and (\ref{eq:7}), we get the following quadratic equation:
	\begin{equation} \label{eq:8}
		M_{j\bar{j}}^2\left( \frac{\alpha'_{j\bar{j}}}{\alpha'_{i\bar{i}}}  \right)^2+\left( M_{i\bar{i}}^2 + M_{j\bar{j}}^2 - 4 M_{i\bar{j}}^2 \right) \left(\frac{\alpha'_{j\bar{j}}}{\alpha'_{i\bar{i}}} \right) + M_{i\bar{i}}^2 = 0 .
	\end{equation}
	
	Solving the above equation, we obtain the solution as follows:
	\begin{equation} \label{eq:11}
		\begin{split}
			\frac{\alpha'_{j\bar{j}}}{\alpha'_{i\bar{i}}} &= \frac{1}{2 M_{j\bar{j}}^2} \biggl[ \bigl(4 M_{i\bar{j}}^2 - M_{i\bar{i}}^2 - M_{j\bar{j}}^2 \bigr) \\
			&\quad \pm \sqrt{\bigl(4 M_{i\bar{j}}^2 - M_{i\bar{i}}^2 - M_{j\bar{j}}^2 \bigr)^2 - 4 M_{i\bar{i}}^2 M_{j\bar{j}}^2} \biggr] 
		\end{split},
	\end{equation}

	By using Eqs (\ref{eq:7}) and (\ref{eq:11}) we can get following equation for $\alpha'_{i\bar{j}}/\alpha'_{j\bar{j}}$ ,
	
	\begin{equation} \label{eq:12}
		\begin{split}
			\frac{\alpha'_{i\bar{j}}}{\alpha'_{j\bar{j}}} &= \frac{1}{4 M_{i\bar{j}}^2} \biggl[ \bigl(4 M_{i\bar{j}}^2 + M_{j\bar{j}}^2 - M_{i\bar{i}}^2 \bigr) \\
			&\quad \pm \sqrt{\bigl(4 M_{i\bar{j}}^2 - M_{i\bar{i}}^2 - M_{j\bar{j}}^2 \bigr)^2 - 4 M_{i\bar{i}}^2 M_{j\bar{j}}^2} \biggr] 
		\end{split}.
	\end{equation}
	Here, we will consider the solutions with the plus sign before the square root term, because solutions with the plus sign give us the values of ratios of slope which are approximately equal to the ratios of experimental values of slopes for some well-established meson multiplets \cite{ref39}. Similarly, for tetraquarks, if we calculate the slope ratios using Eq.~(\ref{eq:31}), the resulting value is found to be closer to the one obtained from the solution with the plus sign, rather than the solution with the minus sign. We have verified this by checking the ratio of $\alpha'_{cc\bar{c}\bar{c}}$ to $\alpha'_{bb\bar{b}\bar{b}}$, using theoretical masses taken from Ref.~\cite{ref105}.
	 So, both of the above equations with the plus sign before square root term can be written as follows:
	
	\begin{equation} \label{eq:9}
		\begin{split}
			\frac{\alpha'_{j\bar{j}}}{\alpha'_{i\bar{i}}} &= \frac{1}{2 M_{j\bar{j}}^2} \biggl[ \bigl(4 M_{i\bar{j}}^2 - M_{i\bar{i}}^2 - M_{j\bar{j}}^2 \bigr) \\
			&\quad + \sqrt{\bigl(4 M_{i\bar{j}}^2 - M_{i\bar{i}}^2 - M_{j\bar{j}}^2 \bigr)^2 - 4 M_{i\bar{i}}^2 M_{j\bar{j}}^2} \biggr] 
		\end{split},
	\end{equation}
	
	\begin{equation} \label{eq:10}
		\begin{split}
			\frac{\alpha'_{i\bar{j}}}{\alpha'_{j\bar{j}}} &= \frac{1}{4 M_{i\bar{j}}^2} \biggl[ \bigl(4 M_{i\bar{j}}^2 + M_{j\bar{j}}^2 - M_{i\bar{i}}^2 \bigr) \\
			&\quad + \sqrt{\bigl(4 M_{i\bar{j}}^2 - M_{i\bar{i}}^2 - M_{j\bar{j}}^2 \bigr)^2 - 4 M_{i\bar{i}}^2 M_{j\bar{j}}^2} \biggr] 
		\end{split}.
	\end{equation}
	These equations give us the important relations between the slope ratios and masses of two body system. Eq. (\ref{eq:9}) can also be expressed in terms of meson masses only, by introducing $k$, where $k$ can be any quark flavor. So, using
	
	\begin{equation} \label{eq:13}
		\frac{\alpha'_{j\bar{j}}}{\alpha'_{i\bar{i}}} = \frac{\alpha'_{k\bar{k}}}{\alpha'_{i\bar{i}}} \times \frac{\alpha'_{j\bar{j}}}{\alpha'_{k\bar{k}}},
	\end{equation}
	we  have,
	
\begin{strip} 
	\begin{align}  \label{eq:14}
		\quad \quad \quad \quad \quad \quad \quad \quad \quad
		&\frac{(4M_{i\bar{j}}^2 - M_{i\bar{i}}^2 - M_{j\bar{j}}^2) + \sqrt{(4M_{i\bar{j}}^2 - M_{i\bar{i}}^2 - M_{j\bar{j}}^2)^2 - 4M_{i\bar{i}}^2 M_{j\bar{j}}^2}}{2M_{j\bar{j}}^2} \notag \\
		&= \frac{((4M_{i\bar{k}}^2 - M_{i\bar{i}}^2 - M_{k\bar{k}}^2) + \sqrt{(4M_{i\bar{k}}^2 - M_{i\bar{i}}^2 - M_{k\bar{k}}^2)^2 - 4M_{i\bar{i}}^2 M_{k\bar{k}}^2})/2M_{k\bar{k}}^2 }{ ((4M_{k\bar{j}}^2 - M_{j\bar{j}}^2 - M_{k\bar{k}}^2) + \sqrt{(4M_{k\bar{j}}^2 - M_{j\bar{j}}^2 - M_{k\bar{k}}^2)^2 - 4M_{j\bar{j}}^2 M_{k\bar{k}}^2})/2M_{k\bar{k}}^2 }
	\end{align}
\end{strip}

If we know all the other masses, we may use this basic relation between meson masses to predict the mass of any meson state. However, due to the lack of experimental and theoretical data in the light-strange tetraquark sector, we will not apply this equation. However, we will establish some inequalities as follows.

	\subsection{\textbf{Linear mass inequalities and quadratic mass inequalities}}
	
	From Eq. (\ref{eq:9}), two noteworthy inequalities can be obtained.  
	
	Since the Regge slopes $\alpha'_{j\bar{j}}$ and $\alpha'_{i\bar{i}}$ must be positive real numbers, their ratio, $\alpha'_{j\bar{j}}/\alpha'_{i\bar{i}}$, should also be a real number. Consequently, from Eq. (\ref{eq:9}), we obtain
	\begin{equation} \label{eq:15}
		|4M^2_{i\bar{j}} - M^2_{i\bar{i}} - M^2_{j\bar{j}}| \geq 2M_{i\bar{i}}M_{j\bar{j}}.
	\end{equation}
	
	When \( i = j \), the inequality \( 4M^2_{i\bar{j}} - M^2_{i\bar{i}} - M^2_{j\bar{j}} \leq 0 \) does not hold. For \( i \neq j \), this inequality is contradicted by the data from well-established meson multiplets. Hence, we conclude that  
	\begin{equation}
		4M^2_{i\bar{j}} - M^2_{i\bar{i}} - M^2_{j\bar{j}} \geq 0.
	\end{equation}  
	As a result, Eq. (\ref{eq:15}) can be rewritten as follows:
	
	\begin{equation} \label{eq:16}
		4M^2_{i\bar{j}} - M^2_{i\bar{i}} - M^2_{j\bar{j}} \geq 2M_{i\bar{i}}M_{j\bar{j}}.
	\end{equation}
	
	By adding \( M^2_{i\bar{i}} \) and \( M^2_{j\bar{j}} \) to both sides, we obtain:
	
	\begin{equation} \label{eq:17}
		2M_{i\bar{j}} \geq M_{i\bar{i}} + M_{j\bar{j}}.
	\end{equation}
	
	If \( i = j \), then \( M_{i\bar{i}} = M_{i\bar{j}} = M_{j\bar{j}} \), which leads to the relation \( 2M_{i\bar{j}} = M_{i\bar{i}} + M_{j\bar{j}} \). 
	
	Conversely, without assuming \( i = j \), if the condition \( 2M_{i\bar{j}} = M_{i\bar{i}} + M_{j\bar{j}} \) holds, then using Eq. (\ref{eq:9}), we can derive the following:
	
	\begin{equation} \label{eq:18}
		\frac{\alpha'_{j\bar{j}}}{\alpha'_{i\bar{i}}} = \frac{M_{i\bar{i}}}{M_{j\bar{j}}}.
	\end{equation}
	
From the derivation of Eq. (\ref{eq:18}), it is evident that this equation holds for mesons within the same multiplet. Given that hadrons on the same Regge trajectory share an identical slope, we obtain  
	\begin{equation} \label{eq:19}
		\frac{\alpha'_{j\bar{j}}}{\alpha'_{i\bar{i}}} = \frac{M_{i\bar{i},J}}{M_{j\bar{j},J}} = \frac{M_{i\bar{i},J+2}}{M_{j\bar{j},J+2}}.
	\end{equation}
	
Using Eq. (\ref{eq:4}), the slopes of specific Regge trajectories can be determined. For $i\bar{i}$ and $j\bar{j}$ mesons, the slopes are expressed as  

\begin{equation} \label{eq:20}
	\alpha'_{i\bar{i}} = \frac{2}{M^2_{i\bar{i},J+2} - M^2_{i\bar{i},J}}, \quad \alpha'_{j\bar{j}} = \frac{2}{M^2_{j\bar{j},J+2} - M^2_{j\bar{j},J}}. 
\end{equation}

Therefore from the above equation we can have,

\begin{equation} \label{eq:21}
	\frac{\alpha'_{j\bar{j}}}{\alpha'_{i\bar{i}}} = \frac{M_{i\bar{i},J+2} + M_{i\bar{i},J}}{M_{j\bar{j},J+2} + M_{j\bar{j},J}} \times \frac{M_{i\bar{i},J+2} - M_{i\bar{i},J}}{M_{j\bar{j},J+2} - M_{j\bar{j},J}}. 
\end{equation}

By merging Eqs. (\ref{eq:19}) and (\ref{eq:21}), we obtain  

\begin{equation} \label{eq:22}
	\frac{\alpha'_{j\bar{j}}}{\alpha'_{i\bar{i}}} = \frac{M_{i\bar{i},J+2} + M_{i\bar{i},J}}{M_{j\bar{j},J+2} + M_{j\bar{j},J}} \times \frac{M_{i\bar{i},J+2} - M_{i\bar{i},J}}{M_{j\bar{j},J+2} - M_{j\bar{j},J}} = \left( \frac{\alpha'_{j\bar{j}}}{\alpha'_{i\bar{i}}} \right)^2. 
\end{equation}

As previously stated, the Regge slope $\alpha'$ is a positive real number. Thus, according to Eq. (\ref{eq:22}), the condition $\alpha'_{j\bar{j}}/\alpha'_{i\bar{i}} = 1$ holds when $2M_{i\bar{j}} = M_{i\bar{i}} + M_{j\bar{j}}$. As a result, using Eq. (\ref{eq:19}), we obtain $M_{i\bar{i},J} = M_{j\bar{j},J}$ and $M_{i\bar{i},J+2} = M_{j\bar{j},J+2}$. This implies $i = j$, given that the $i\bar{i}$ and $j\bar{j}$ states share the same $J^P$.  

Based on the preceding analysis, it follows that the equation $2M_{i\bar{j}} = M_{i\bar{i}} + M_{j\bar{j}}$ holds if and only if $i = j$. Consequently, for the case where $i \neq j$, from (\ref{eq:17}) we obtain  

\begin{equation} \label{eq:23}
	2M_{i\bar{j}} > M_{i\bar{i}} + M_{j\bar{j}}. 
\end{equation}

Studies have suggested that the slopes of Regge trajectories decrease as the quark mass increases \cite{ref44,ref45,ref100,ref101,ref102,ref103,ref104,ref47,ref53}. As a result, $\alpha'_{j\bar{j}}/\alpha'_{i\bar{i}} < 1$ when the mass of the $j$ quark is greater than that of the $i$ quark. Hence, from Eq. (\ref{eq:9}), one can derive

\begin{equation} \label{eq:24}
	\begin{aligned}
		\frac{1}{2M_{j\bar{j}}^{2}} \Bigg[ &\left( 4M_{i\bar{j}}^{2} - M_{i\bar{i}}^{2} - M_{j\bar{j}}^{2} \right) \\
		&+ \sqrt{\left( 4M_{i\bar{j}}^{2} - M_{i\bar{i}}^{2} - M_{j\bar{j}}^{2} \right)^{2} - 4M_{i\bar{i}}^{2} M_{j\bar{j}}^{2}} \Bigg] < 1
	\end{aligned}
\end{equation}

Since the square root quantity in the above equation is positive, we can have

	\begin{equation} \label{eq:25}
		2M_{j\bar{j}}^2 - (4M_{i\bar{j}}^2 - M_{i\bar{i}}^2 - M_{j\bar{j}}^2) > 0
	\end{equation}

Using Eqs. (\ref{eq:24}) and (\ref{eq:25}),	
	
\begin{equation} \label{eq:26}
	\begin{aligned}
		(4M_{i\bar{j}}^2 - M_{i\bar{i}}^2 - M_{j\bar{j}}^2)^2 
		&- 4M_{i\bar{i}}^2 M_{j\bar{j}}^2 \\
		&< \left[ 2M_{j\bar{j}}^2 - (4M_{i\bar{j}}^2 - M_{i\bar{i}}^2 - M_{j\bar{j}}^2) \right]^2
	\end{aligned}
\end{equation}

The last two inequalities can be simplified to,
\begin{equation} \label{eq:27}
	2M_{i\bar{j}}^2 < M_{i\bar{i}}^2 + M_{j\bar{j}}^2
\end{equation}

Using, Eqs. (\ref{eq:23}) and (\ref{eq:27}), we can get the following constraint relation for $M_{i\bar{j}}$.
\begin{equation} \label{eq:28}
	\frac{M_{i\bar{i}} + M_{j\bar{j}}}{2} < M_{i\bar{j}} < \sqrt{\frac{M_{i\bar{i}}^2 + M_{j\bar{j}}^2}{2}}
\end{equation}

The mass inequality presented above provides both the upper and lower bounds for the mass of the $M_{i\bar{j}}$ meson. We will apply this relationship to estimate the mass range of the yet-to-be-discovered tetraquarks in the following section.

	\section{Mass Spectra of Tetraquark}
	\subsection{The four-quark state in diquark-antidiquark picture}
	Here, we calculate the mass spectra of light-light, strange-strange and light-strange tetraquarks while considering them as the bound states of two clusters (diquark and anti-diquark). The diquarks are thought of as two coupled quarks free of any interior spatial excitation. The diquark can only be discovered contained within hadrons and employed as an effective degree of freedom since a pair of quarks cannot be a color singlet. A tetraquark in a color singlet state can arise from two distinct diquark-antidiquark pairings: (i) a color anti-triplet diquark paired with a color triplet anti-diquark, or (ii) a color sextet diquark paired with a color anti-sextet anti diquark. 
	So, if we consider tetraquark as two body system made up of diquark and anti-diquark then we can use equation (\ref{eq:28}), to get the intervals for tetraquark masses.

	\subsection{\textbf{Mass Spectra of $qq\bar{q}\bar{q}$, \(ss\bar{s}\bar{s}\) and  $qq\bar{s}\bar{s}$ tetraquarks in the ($J,M^{2}$) plane}}
	
	In this work we use equation (\ref{eq:28}) to evaluate the mass ranges of ground-state of the $qq\bar{s}\bar{s}$ tetraquark. $qq\bar{s}\bar{s}$  tetraquark is considered to be composed of $qq$ diquark and $\bar{s}\bar{s}$ anti-diquark. Here, $q$ is up or down quark ($q=u/d$). In eq (\ref{eq:28}), if we take $i=[qq]$, $j=[ss]$, we get the following relation:
	
	\begin{equation} \label{eq:29}
		\frac{M_{qq\bar{q}\bar{q}} + M_{ss\bar{s}\bar{s}}}{2} < M_{qq\bar{s}\bar{s}} < \sqrt{\frac{M_{qq\bar{q}\bar{q}}^2 + M_{ss\bar{s}\bar{s}}^2}{2}}
	\end{equation}
	
	In this work, the input masses for the \(qq\bar q\bar q\) and \(ss\bar s\bar s\) systems are taken from Refs.~\cite{ref54,ref55}, respectively. As no experimental determinations exist for these sectors, we rely on theoretical predictions. For completeness, we consulted the Review of Particle Physics (2024) and the PDG topical reviews on light-meson spectroscopy and scalar mesons \cite{ref60}. These sources emphasize that the light scalar nonet has a debated internal structure - being alternatively interpreted as ordinary \(q\bar q\), meson--meson molecular, tetraquark, or mixed states. Importantly, the PDG does not list any purely light or fully-strange states with firm tetraquark assignments and well-defined \(J^{PC}\) suitable for a PDG-only fitting. Consequently, the trajectories for the purely light (\(qq\bar q\bar q\)) and fully-strange (\(ss\bar s\bar s\)) sectors are constructed using the theoretical inputs of Refs.~\cite{ref54,ref55}. Since the adopted inputs are quoted as central theoretical values without associated statistical uncertainties, a formal \(\chi^2\) analysis yielding error bars is not feasible in these cases.

	By substituting the masses of $qq\bar{q}\bar{q}$ and $ss\bar{s}\bar{s}$ (for $J^P = 0^+$, $1^+$, $2^+$), we derive the range of ground-state masses for the $qq\bar{s}\bar{s}$ tetraquark, which are 1.862-1.911 GeV for $J^P = 0^+$, 2.073-2.088 GeV for $J^P = 1^+$, and 2.157-2.168 GeV for $J^P = 2^+$.

	
	To find the higher excited states, the Regge slopes for these tetraquarks are calculated. For example, to determine the value of $\alpha'$ for the $qq\bar{s}\bar{s}$ tetraquark system, we apply relation (\ref{eq:10}). By substituting the appropriate values of $i$ and $j$ and solving for $\alpha'_{qq\bar{s}\bar{s}}$, we obtain the following expression:

	\begin{equation} \label{eq:30}
		\begin{aligned}
			\alpha'_{qq\bar{s}\bar{s}} &= \frac{\alpha'_{ss\bar{s}\bar{s}}}{4 M_{qq\bar{s}\bar{s}}^2} \Biggl[ \Bigl(4 M_{qq\bar{s}\bar{s}}^2 + M_{ss\bar{s}\bar{s}}^2 - M_{qq\bar{q}\bar{q}}^2 \Bigr) \\
			&\quad + \sqrt{\Bigl(4 M_{qq\bar{s}\bar{s}}^2 - M_{qq\bar{q}\bar{q}}^2 - M_{ss\bar{s}\bar{s}}^2 \Bigr)^2 - 4 M_{qq\bar{q}\bar{q}}^2 M_{ss\bar{s}\bar{s}}^2} \Biggr].
		\end{aligned}
	\end{equation}
	 
	Now, we can determine the slope of the Regge trajectory for \(ss\bar{s}\bar{s}\) tetraquark using equation (\ref{eq:31}),
	
	\begin{equation} \label{eq:32}
		\alpha'_{ss\bar{s}\bar{s}} = \frac{1}{M_{ss\bar{s}\bar{s}(1^-)}^2 - M_{ss\bar{s}\bar{s}(0^+)}^2} .
	\end{equation} 
	The value of \(M_{ss\bar{s}\bar{s}(1^-)}\) is also taken from Ref.~\cite{ref55} for the calculation of the slope of the $ss\bar{s}\bar{s}$ tetraquark. The calculated Regge slopes for various \(J^P\) values of $ss\bar{s}\bar{s}$ are presented in Table \ref{tab:Slope}. By using the value of \(\alpha'_{ss\bar{s}\bar{s}}\) and Eq. (\ref{eq:31}), we can compute the masses of the excited states of the $ss\bar{s}\bar{s}$ tetraquark, which are listed in Table \ref{tab:mass_spectrassss}.

By substituting the values of \(M_{qq\bar{q}\bar{q}}\), \(M_{ss\bar{s}\bar{s}}\), and \(\alpha'_{ss\bar{s}\bar{s}}\) into Eq. (\ref{eq:30}), we can express \(\alpha'_{qq\bar{s}\bar{s}}\) as a function of \(M_{qq\bar{s}\bar{s}}\). Over the interval (1.862-1.911), this function is increasing. The resulting range for \(\alpha'_{qq\bar{s}\bar{s}}\) for \(J^P=0^+\) is between 0.72840 and 0.85113, which is summarized in Table \ref{tab:Slope}. For other \(J^P\) values, the corresponding ranges of \(\alpha'_{qq\bar{s}\bar{s}}\) are also provided in Table \ref{tab:Slope}.

	In the similar manner, by using equation (\ref{eq:6}) and (\ref{eq:30}) we can get the below equation:
	\\
	\\
	
	
	
	\begin{strip}
\begin{equation} \label{eq:33}
	\alpha'_{qq\bar{q}\bar{q}} = \frac{1}{
		\left(
		\frac{2}{
			\alpha'_{ss\bar{s}\bar{s}} \cdot \frac{1}{4M_{qq\bar{s}\bar{s}}^{2}}
			\left(
			\left(4M_{qq\bar{s}\bar{s}}^{2} + M_{ss\bar{s}\bar{s}}^{2} - M_{qq\bar{q}\bar{q}}^{2}\right) 
			+ \sqrt{
				\left(4M_{qq\bar{s}\bar{s}}^{2} - M_{qq\bar{q}\bar{q}}^{2} - M_{ss\bar{s}\bar{s}}^{2}\right)^{2} 
				- 4M_{qq\bar{q}\bar{q}}^{2}M_{ss\bar{s}\bar{s}}^{2}
			}
			\right)
		}
		- \frac{1}{\alpha'_{ss\bar{s}\bar{s}}}
		\right)
	}
\end{equation}
		\end{strip}

In the equation above, by substituting the values of \(M_{qq\bar{q}\bar{q}}\), \(M_{ss\bar{s}\bar{s}}\), and \(\alpha'_{ss\bar{s}\bar{s}}\), we can express \(\alpha'_{qq\bar{q}\bar{q}}\) as a function of \(M_{qq\bar{s}\bar{s}}\). Over the interval (1.862-1.911), this function for \(\alpha'_{qq\bar{q}\bar{q}}\) is increasing, and the resulting range for \(\alpha'_{qq\bar{q}\bar{q}}\) for \(J^P=0^+\) is between 0.94779 and 1.51704, as shown in Table \ref{tab:Slope}. The range of \(\alpha'_{qq\bar{q}\bar{q}}\) for other \(J^P\) values is also provided in Table \ref{tab:Slope}.


	Further, by Eq. (\ref{eq:4}), we can write mass of $qq\bar{s}\bar{s}$ tetraquark for excited state as following:
	\begin{equation} \label{eq:34}
		M_{J+k(qq\bar{s}\bar{s})} = \sqrt{M_{J(qq\bar{s}\bar{s})}^2 + \frac{k}{\alpha'_{qq\bar{s}\bar{s}}}} ,
	\end{equation}
	where, k is an integer number.
	By utilizing Eqs. (\ref{eq:34}) and (\ref{eq:30}), we can express \(M_{J+k(qq\bar{s}\bar{s})}\) as a function of \(M_{J(qq\bar{s}\bar{s})}\). For instance, for \(J^P = 0^+\) and \(k = 1\), the function over the interval (1.862-1.911) yields a minimum value of 2.185 and a maximum value of 2.200 for \(M_{J+k(qq\bar{s}\bar{s})}\) with \(J^P = 1^-\). Therefore, the range of (2.185-2.200) is obtained for the \(qq\bar{s}\bar{s}\) tetraquark with \(J^P = 1^-\), as shown in Table \ref{tab:mass_spectraqqss}. Similarly, the excited states for the \(qq\bar{s}\bar{s}\) tetraquark are also computed and listed in Table \ref{tab:mass_spectraqqss}.
	\begin{table}[h]
		\caption{Values of Regge Slopes for \(qq\bar{q}\bar{q}\), \(ss\bar{s}\bar{s}\) and \(qq\bar{s}\bar{s}\) tetraquarks in \((J,M^2)\) plane (in \(\text{GeV}^{-2}\))}{\label{tab:Slope}}
		\begin{tabular}{cccc}
			\toprule
			\(J^P\) & \(\alpha'_{qq\bar{q}\bar{q}} \, (\text{GeV}^{-2})\) & \(\alpha'_{ss\bar{s}\bar{s}} \, (\text{GeV}^{-2})\) & \(\alpha'_{qq\bar{s}\bar{s}} \, (\text{GeV}^{-2})\) \\
			\midrule
			\(0^+\) & 0.94779-1.51704 & 0.59149 & 0.72840-0.85113 \\
			\(1^+\) & 0.72874-0.92847 & 0.57189 & 0.64086-0.70785 \\
			\(2^+\) & 0.70671-0.86573 & 0.57535 & 0.63430-0.69202 \\
			\bottomrule
		\end{tabular}
	\end{table}

In a similar manner, we can derive the corresponding formula for the \(qq\bar{q}\bar{q}\) tetraquark using Eq. (\ref{eq:4}).

	\begin{equation} \label{eq:35}
		M_{J+k(qq\bar{q}\bar{q})} = \sqrt{M_{J(qq\bar{q}\bar{q})}^2 + \frac{k}{\alpha'_{qq\bar{q}\bar{q}}}} ,
	\end{equation}
	
	By applying Eqs. (\ref{eq:35}) and (\ref{eq:33}), we can express \( M_{J+k(qq\bar{q}\bar{q})} \) as a function of \( M_{J(qq\bar{q}\bar{q})} \). Using this approach, we have determined the range for the masses of other excited states of the \( qq\bar{q}\bar{q} \) tetraquark, which are presented in Table \ref{tab:mass_spectraqqqq}.

	 The estimated mass spectra of $qq\bar{q}\bar{q}$, $ss\bar{s}\bar{s}$ and $qq\bar{s}\bar{s}$ systems are presented in Tables \ref{tab:mass_spectraqqqq}, \ref{tab:mass_spectrassss} and \ref{tab:mass_spectraqqss} respectively as stated above. Here, we have also compared our results with the two-meson threshold value. Also, the possible resonance states are mentioned with their experimental masses and \(J^P\) values. These resonance states are taken from latest Particle Data Group (PDG) \cite{ref60}.

	 \subsection{\textbf{Statistical comparison of predicted intervals and experimental masses}}
	 When the Regge analysis yields only a predicted mass interval $[a,b]$ (rather than a Gaussian error about a central value) we model the theoretical prediction as a uniform distribution on $[a,b]$. This is a non-informative, reproducible choice in the absence of further information on the distribution of the theoretical prediction inside the interval. From the interval we define the nominal predicted mass (midpoint)
	
	 \begin{equation}
	 	m=\frac{a+b}{2}.
	 \end{equation}
	 
	 The theoretical uncertainty corresponding to a uniform distribution is  
	 \begin{equation}
	 	\sigma_{\rm th}=\frac{b-a}{\sqrt{12}}.
	 \end{equation}
	 
	 The experimental uncertainty $\sigma_{\rm exp}$ is taken from the Particle Data Group (PDG) entry for the resonance; if PDG reports asymmetric errors we conservatively use the larger side. The combined uncertainty is then  
	 \begin{equation}
	 	\sigma=\sqrt{\sigma_{\rm th}^2+\sigma_{\rm exp}^2}.
	 \end{equation}
	 
	 Finally, we quantify the compatibility between the theoretical prediction and an experimental mass $M_{\rm exp}$ by the z-score  
	 \begin{equation}
	 	z=\frac{|M_{\rm exp}-m|}{\sigma}.
	 \end{equation}
	 
	 and adopt the following qualitative classification: $z\le 1$ (Strongly compatible), $1<z\le 2$ (Plausible), $2<z\le 3$ (Weak), and $z>3$ (Incompatible). 
	 All z-scores and compatibility are reported in the comparison tables; claims of ``compatibility'' in the text refer to the numerical z-score classification above.
	 
	 We illustrate this procedure for the state $1^1P_1$ of \(qq\bar{q}\bar{q}\) tetraquark and the PDG resonance $X(1650)$:
	 
	 The predicted interval is 
	 	\([a,b]=[1.645,\;1.761]\ \mathrm{GeV}\).
	 
The midpoint (nominal prediction) will be  
\[
m=\frac{a+b}{2}=1.703\ \mathrm{GeV}.
\]

The corresponding uniform theoretical uncertainty is  
\[
\sigma_{\rm th}=\frac{b-a}{\sqrt{12}}=0.033486\ \mathrm{GeV}.
\]

For comparison, the experimental mass and uncertainty from the PDG are  
\[
M_{\rm exp}\pm \sigma_{exp}=1.652\pm 0.007\ \mathrm{GeV}.
\]
Here we adopt the quoted symmetric error; in the case of asymmetric errors, the larger side is used.  

The combined uncertainty is then  
\[
\sigma=\sqrt{\sigma_{\rm th}^2+\sigma_{\rm exp}^2}=0.034210\ \mathrm{GeV}.
\]

This yields a z-score  
\[
z=\frac{|M_{\rm exp}-m|}{\sigma}=1.491.
\]

	 According to the adopted thresholds, this corresponds to the verdict: Plausible.

	 This worked example demonstrates the numeric steps used throughout the paper. For reproducibility, the comparison table for \(qq\bar{q}\bar{q}\) tetraquark, list $m$, $\sigma_{\rm th}$, the experimental $\sigma$ used, the combined $\sigma$, the z-score and the qualitative verdict are given in Table \ref{tabqqqqn}
	 
	 Z-scores and Verdicts for \(qq\bar{s}\bar{s}\) tetraquarks is embedded in the Table.~\ref{tab:mass_spectraqqss}.

	



	\begin{table*}[htbp]
		\centering
		\caption{Mass spectra of $qq\bar{q}\bar{q}$ tetraquark (in GeV)}
		\begin{tabular}{lccccccc}
			\toprule
			State & \parbox{0.5 cm}{$J^{PC}$} & \parbox{2 cm}{$M_{(\text{calc})}$ \\ Calculated Mass \\ (in GeV)} & \parbox{2cm}{Meson Threshold} & \parbox{2 cm}{$M_{\text{th}}$ \\ (Threshold mass)} & \parbox{3.4 cm}{Resonance \\ (Experimental 
				$J^{PC}$ value) (PDG) \cite{ref60}} & 
			\parbox{3.4cm}{Resonance State Mass \\ (Experimental) \\ (in GeV) [PDG] \cite{ref60}} \\
			\midrule
			$1^1 P_1$ & $1^{--}$ & 1.645-1.761 & $h_1(1170) \pi^+$ & 1.306 & $X(1650)$ $(J^{PC}=?^{?-})$ & $1.652 \pm 0.007$ \\
			\cmidrule(lr){1-7}
			$1^1 D_2$ & $2^{++}$ & 1.834-2.039 & $\rho(770) \rho(770)$ & 1.550 & 
			\begin{tabular}[c]{@{}l@{}}
				$a_2(1950)$ $(J^{PC}=2^{++})$ \\
				$f_2(2000)$ $(J^{PC}=2^{++})$ \\
				$a_2(2030)$ $(J^{PC}=2^{++})$ 
			\end{tabular} & 
			\begin{tabular}[c]{@{}l@{}}
				$1.950 ^{+0.030}_{-0.070}$ \\
				$2.001 \pm 0.010$ \\
				$2.030 \pm 0.020$
			\end{tabular} \\
			\cmidrule(lr){1-7}
			$1^1 F_3$ & $3^{--}$ & 2.006-2.283 & $\rho_3 (1690) f_0 (500)$ & 2.203 & 
			\begin{tabular}[c]{@{}l@{}}
				$\omega_3(2255)$ $(J^{PC}=3^{--})$ \\
			\end{tabular} & 
			\begin{tabular}[c]{@{}l@{}}
				$2.255 \pm 0.015$ \\
			\end{tabular} \\
			\cmidrule(lr){1-7}
			$1^1 G_4$ & $4^{++}$ & 2.164-2.504 & $f_2 (1270) f_2 (1270)$ & 2.550 & \begin{tabular}[c]{@{}l@{}}
				$a_4(2255)$ $(J^{PC}=4^{++})$ \\
			\end{tabular} & 
			\begin{tabular}[c]{@{}l@{}}
				$2.237 \pm 0.005$ \\
			\end{tabular} \\
			\cmidrule(lr){1-7}
			$1^3 P_2$ & $2^{-+}$ & 2.098-2.167 & $\rho(770) h_1(1170)$ & 1.941 & \begin{tabular}[c]{@{}l@{}}
			\end{tabular} & 
			\begin{tabular}[c]{@{}l@{}}
			\end{tabular} \\
			\cmidrule(lr){1-7}
			$1^3 D_3$ & $3^{+-}$ & 2.340-2.463 & $f_2(1270) h_1(1170)$ & 2.441 & \begin{tabular}[c]{@{}l@{}}
				$X(2340)$ $(J^{PC}=?^{??})$ \\
			\end{tabular} & 
			\begin{tabular}[c]{@{}l@{}}
				$2.340 \pm 0.020$ \\
			\end{tabular} \\
			\cmidrule(lr){1-7}
			$1^3 F_4$ & $4^{-+}$ & 2.560-2.728 & $a_4 (1970) \eta$ & 2.515 & \begin{tabular}[c]{@{}l@{}}
				$X(2600)$ $(J^{PC}=?^{??})$ \\
				$X(2632)$ $(J^{PC}=?^{??})$ \\
				$X(2680)$ $(J^{PC}=?^{??})$ \\
			\end{tabular} & 
			\begin{tabular}[c]{@{}l@{}}
				$2.618 \pm 0.002$ \\
				$2.635 \pm 0.003$ \\			
				$2.676 \pm 0.027$ \\
			\end{tabular} \\
			\cmidrule(lr){1-7}
			$1^3 G_5$ & $5^{+-}$ & 2.762-2.968 & $\rho_5 (2350) \eta$ & 2.878 & & \\
			\cmidrule(lr){1-7}
			$1^5 P_3$ & $3^{--}$ & 2.214-2.272 & $f_2(1270) \rho(770)$ & 2.051 & \begin{tabular}[c]{@{}l@{}}
				$\omega_3(2255)$ $(J^{PC}=3^{--})$ \\
			\end{tabular} & 
			\begin{tabular}[c]{@{}l@{}}
				$2.255 \pm 0.015$ \\
			\end{tabular} \\
			\cmidrule(lr){1-7}
			$1^5 D_4$ & $4^{++}$ & 2.460-2.565 & $\omega_3(1670) \rho(770)$ & 2.442 & & \\
			\cmidrule(lr){1-7}
			$1^5 F_5$ & $5^{--}$ & 2.684-2.827 & $a_4 (1970) \rho(770)$ & 2.742 & & \\
			\cmidrule(lr){1-7}
			$1^5 G_6$ & $6^{++}$ & 2.891-3.067 & $f_0 (2510) f_0 (500)$ & 2.983 & \begin{tabular}[c]{@{}l@{}}
				$f_0(3100)$ $(J^{PC}=6^{++})$ \\
			\end{tabular} & 
			\begin{tabular}[c]{@{}l@{}}
				$3.100 \pm 0.100$ \\
			\end{tabular} \\
			\bottomrule
		\end{tabular}
		\label{tab:mass_spectraqqqq}
	\end{table*}

\begin{table*}[htbp]
	\centering
	\caption{\textbf{Results for predicted masses and z-scores of $qq\bar{q}\bar{q}$ tetraquark states compared with PDG data.}}
	\label{tabqqqqn}
	\begin{tabular}{l l l l c c c c c c l}
		\toprule
		State & $J^{PC}$ & \parbox{1.8 cm}{Resonance \\ (Experimental 
			$J^{PC}$ value) (PDG) \cite{ref60} }  & \parbox{2.0 cm}{$M_{\mathrm{exp}}\pm \sigma_{exp}$ \\(GeV) \cite{ref60} }& 
		$a$ & $b$ & $m$ & $\sigma_{\mathrm{th}}$ & $\sigma$ & $z$ & Verdict \\
		\midrule
		$1^1 P_1$ & $1^{--}$ & $X(1650)$ $(?^{?-})$ & $1.652 \pm 0.007$ & 1.645 & 1.761 & 1.703 & 0.033486 & 0.03421 & 1.490786 & Plausible \\
		\midrule
		\multirow{3}{*}{$1^1 D_2$} & \multirow{3}{*}{$2^{++}$} & $a_2(1950)$ $(2^{++})$ & $1.950^{+0.030}_{-0.070}$ & \multirow{3}{*}{1.834} & \multirow{3}{*}{2.039} & \multirow{3}{*}{1.9365} & \multirow{3}{*}{0.059178} & 0.091663 & 0.147279 & Strongly compatible \\
		& & $f_2(2000)$ $(2^{++})$ & $2.001 \pm 0.010$ &  &  &  &  & 0.060017 & 1.074689 & Plausible \\
		& & $a_2(2030)$ $(2^{++})$ & $2.030 \pm 0.020$ &  &  &  &  & 0.062467 & 1.496799 & Plausible \\
		\midrule
		$1^1 F_3$ & $3^{--}$ & $\omega_3(2255)$ $(3^{--})$ & $2.255 \pm 0.015$ & 2.006 & 2.283 & 2.1445 & 0.079963 & 0.081358 & 1.358199 & Plausible \\
		\midrule
		$1^1 G_4$ & $4^{++}$ & $a_4(2255)$ $(4^{++})$ & $2.237 \pm 0.005$ & 2.164 & 2.504 & 2.334 & 0.09815 & 0.098277 & 0.987008 & Strongly compatible \\
		\midrule
		$1^3 P_2$ & $2^{-+}$ & -- & -- & 2.098 & 2.167 & 2.1325 & 0.019919 & -- & -- & -- \\
		\midrule
		$1^3 D_3$ & $3^{+-}$ & $X(2340)$ $(?^{??})$ & $2.340 \pm 0.020$ & 2.34 & 2.463 & 2.4015 & 0.035507 & 0.040752 & 1.509117 & Plausible \\
		\midrule
		\multirow{3}{*}{$1^3 F_4$} & \multirow{3}{*}{$4^{-+}$} & $X(2600)$ $(?^{??})$ & $2.618 \pm 0.002$ & \multirow{3}{*}{2.56} & \multirow{3}{*}{2.728} & \multirow{3}{*}{2.644} & \multirow{3}{*}{0.048497} & 0.048539 & 0.535656 & Strongly compatible \\
		& & $X(2632)$ $(?^{??})$ & $2.635 \pm 0.003$ &  &  &  &  & 0.04859 & 0.185223 & Strongly compatible \\
		& & $X(2680)$ $(?^{??})$ & $2.676 \pm 0.027$ &  &  &  &  & 0.055507 & 0.576506 & Strongly compatible \\
		\midrule
		$1^3 G_5$ & $5^{+-}$ & -- & -- & 2.762 & 2.968 & 2.865 & 0.059467 & -- & -- & -- \\
		\midrule
		$1^5 P_3$ & $3^{--}$ & $\omega_3(2255)$ $(3^{--})$ & $2.255 \pm 0.015$ & 2.214 & 2.272 & 2.243 & 0.016743 & 0.02248 & 0.533817 & Strongly compatible \\
		\midrule
		$1^5 D_4$ & $4^{++}$ & -- & -- & 2.46 & 2.565 & 2.5125 & 0.030311 & -- & -- & -- \\
		\midrule
		$1^5 F_5$ & $5^{--}$ & -- & -- & 2.684 & 2.827 & 2.7555 & 0.041281 & -- & -- & -- \\
		\midrule
		$1^5 G_6$ & $6^{++}$ & $f_0(3100)$ $(6^{++})$ & $3.100 \pm 0.100$ & 2.891 & 3.067 & 2.979 & 0.050807 & 0.112167 & 1.078753 & Plausible \\
		\bottomrule
	\end{tabular}
\end{table*}

	\begin{table*}[htbp]
		\centering
		\caption{Mass spectra of $ss\bar{s}\bar{s}$ tetraquark (in GeV)}
		\begin{tabular}{lcccccc}
			\toprule
			State & \parbox{0.5 cm}{$J^P$} & \parbox{2 cm}{$M_{(\text{calc})}$ \\ Calculated Mass \\ (in GeV)} & \parbox{2cm}{Meson Threshold} & \parbox{2 cm}{$M_{\text{th}}$ \\ (Threshold mass)} & \parbox{3.4 cm}{Resonance \\ (Experimental 
				$J^P$ value) (PDG) \cite{ref60}} & 
			\parbox{3.4cm}{Resonance State Mass \\ (Experimental) \\ (in GeV) [PDG] \cite{ref60}} \\
			\midrule
			$1^1 D_2$ & $2^+$ & 2.939 & $h_1 (1411) f_1 (1510)$ & 2.927 & & \\
			$1^1 F_3$ & $3^-$ & 3.214 & $\pi_2 (1670) f_1 (1510)$ & 3.188 & & \\
			$1^1 G_4$ & $4^+$ & 3.467 & $\phi_3 (1850) \rho(1570)$ & 3.424 & & \\
			$1^3 D_3$ & $3^+$ & 2.982 & $f_2 (1270) a_1 (1640)$ & 2.931 & & \\
			$1^3 F_4$ & $4^-$ & 3.262 & $\rho_3 (1990) a_1 (1260)$ & 3.242 & & \\
			$1^3 G_5$ & $5^+$ & 3.520 & $\phi_3 (1850) \eta_2 (1645)$ & 3.471 & & \\
			$1^5 D_4$ & $4^+$ & 2.982 & $\phi_3 (1850) \phi(1020)$ & 2.873 & & \\
			$1^5 F_5$ & $5^-$ & 3.262 & $f_J (2220) \phi(1020)$ & 3.250 & & \\
			$1^5 G_6$ & $6^+$ & 3.520 & $f_J (2220) f_2 (1270)$ & 3.506 & & \\
			\bottomrule
		\end{tabular}
		\label{tab:mass_spectrassss}
	\end{table*}

	
	\begin{table*}[htbp]
		\centering
		\caption{\textbf{Mass spectra of $qq\bar{s}\bar{s}$ tetraquark (in GeV)}}
		\begin{tabular}{l c c c c l l c l}
			\toprule
			State & \parbox{0.5 cm}{$J^{PC}$} & \parbox{2 cm}{$M_{(\text{calc})}$ \\ Calculated Mass \\ (in GeV)} & \parbox{2cm}{Meson Threshold} & \parbox{2 cm}{$M_{\text{th}}$ \\ (Threshold mass)} & \parbox{2.0 cm}{Resonance \\ (Experimental 
				$J^{PC}$ value) (PDG) \cite{ref60}} & 
			\parbox{2.0 cm}{$M_{\mathrm{exp}}\pm \sigma_{exp}$ \\(GeV) \cite{ref60} } & $z$ & Verdict \\
			\midrule
			$1^1 S_0$ & $0^{++}$ & 1.862--1.911 & $K^+ K^+$ & 0.987 &  &  & -- & -- \\
			$1^1 P_1$ & $1^{--}$ & 2.185--2.200 & $K_1(1270) K^+$ & 1.747 &  &  & -- & -- \\
			$1^1 D_2$ & $2^{++}$ & 2.449--2.493 & $K^* (892) K^* (892)$ & 1.783 &  &  & -- & -- \\
			$1^1 F_3$ & $3^{--}$ & 2.679--2.754 & $K^* (892) K_2^* (1430)$ & 2.317 &  &  & -- & -- \\
			$1^1 G_4$ & $4^{++}$ & 2.890--2.993 & $K_4^* (2045) K_0^* (700)$ & 2.893 &  &  & -- & -- \\
			$1^3 S_1$ & $1^{+-}$ & 2.073--2.088 & $K^* (892) K^+$ & 1.385 & $X(2075)$  $(1^{+?})$ & $2.084^{+0.004}_{-0.002}$ & 0.594 & Strongly compatible \\
			$1^3 P_2$ & $2^{-+}$ & 2.403--2.420 & $K^* (892) K_1 (1270)$ & 2.145 &  &  & -- & -- \\
			$1^3 D_3$ & $3^{+-}$ & 2.680--2.724 & $K_2^* (1430) K_1 (1270)$ & 2.680 &  &  & -- & -- \\
			$1^3 F_4$ & $4^{-+}$ & 2.932--2.996 & $K_2^* (1430) K_2 (1580)$ & 3.005 &  &  & -- & -- \\
			$1^3 G_5$ & $5^{+-}$ & 3.164--3.246 & $K_1 (1270) K_4^* (2045)$ & 3.315 & $X(3250)$ $(?^{??})$ & $3.245 \pm 0.008$ & 1.601 & Plausible \\
			$1^5 S_2$ & $2^{++}$ & 2.157--2.168 & $K_1 (1770) K^+$ & 2.267 & \begin{tabular}[c]{@{}l@{}}
				$a_2(2175)$ $(2^{++})$
			\end{tabular} & 
			\begin{tabular}[c]{@{}l@{}}
				$2.175 \pm 0.040$
			\end{tabular} & 0.312 & Strongly compatible \\
			$1^5 P_3$ & $3^{--}$ & 2.479--2.496 & $K_2^* (1430) K^* (892)$ & 2.319 &  &  & -- & -- \\
			$1^5 D_4$ & $4^{++}$ & 2.755--2.793 & $K_3^* (1780) K^* (892)$ & 2.671 &  &  & -- & -- \\
			$1^5 F_5$ & $5^{--}$ & 3.006--3.063 & $K_4^* (2045) K^* (892)$ & 2.940 &  &  & -- & -- \\
			$1^5 G_6$ & $6^{++}$ & 3.238--3.310 & $K_4^* (2380) K^* (892)$ & 3.274 & \begin{tabular}[c]{@{}l@{}}
				$X(3250)$ $(?^{??})$
			\end{tabular} & 
			\begin{tabular}[c]{@{}l@{}}
				$3.250 \pm 0.009$
			\end{tabular} & 1.060 & Plausible \\
			\bottomrule
		\end{tabular}
		\label{tab:mass_spectraqqss}
	\end{table*}

	\subsection{\textbf{Mass Spectra of $qq\bar{q}\bar{q}$, $ss\bar{s}\bar{s}$ and $qq\bar{s}\bar{s}$ tetraquarks in the $(n,M^2)$ plane}}
	
	In this section, we will determine Regge parameters for $qq\bar{q}\bar{q}$, $ss\bar{s}\bar{s}$ and \(qq\bar{s}\bar{s}\) tetraquarks in the \((n,M^2)\) plane to evaluate masses of radial excited states. The general linear equation of Regge trajectory in \((n,M^2)\) plane can be written as,
	\begin{equation} \label{eq:36}
		n = \beta(M) = \beta(0) + \beta' M^2 ,
	\end{equation}
	where \(n = 1, 2, 3, \ldots\) is the radial principal quantum number, and \(\beta(0)\) and \(\beta'\) are the intercept and slope of the  trajectory in the \((n,M^2)\) plane. The Regge parameters are assumed to be the same for all tetraquark multiplets lying on the same Regge line. 
	Here also, we have utilized a similar method to determine the Regge parameters as we have previously applied in the ($J,M^{2}$) plane.	
	
	From Eq. (\ref{eq:36}), we can determine the slope by the following equation 
	\begin{equation} \label{eq:37}
		\beta' = \frac{1}{M_{ss\bar{s}\bar{s}(2S)}^2 - M_{ss\bar{s}\bar{s}(1S)}^2}, 
	\end{equation}
	Since experimental data is not available, we have also taken the masses of the \( 1^1 S_0 \) and \( 2^1 S_0 \) states of the all-strange (\( ss\bar{s}\bar{s} \)) tetraquark from Ref.~\cite{ref55} for our calculations. From Eq. (\ref{eq:37}), we obtain \( \beta'_{ss\bar{s}\bar{s}} = 0.30704 \, \text{GeV}^{-2} \), which is listed in Table \ref{tab:SlopenM2} along with the slopes for other \( J^P \) values. Using these values, the excited states in the \( (n,M^2) \) plane for the \( ss\bar{s}\bar{s} \) tetraquark are computed and presented in Table \ref{tab:mass_spectra_ssssn}.
	
	We assume that Eqs. (\ref{eq:5}) and (\ref{eq:6}), which hold in the \( (J, M^2) \) plane, are also applicable in the \( (n, M^2) \) plane. Consequently, from Eq. (\ref{eq:30}), we can get the following equation in the \( (n, M^2) \) plane:
	
	\begin{equation} \label{eq:38}
		\begin{aligned}
			\beta'_{qq\bar{s}\bar{s}} &= \frac{\beta'_{ss\bar{s}\bar{s}}}{4 M_{qq\bar{s}\bar{s}}^2} \Biggl[ \Bigl(4 M_{qq\bar{s}\bar{s}}^2 + M_{ss\bar{s}\bar{s}}^2 - M_{qq\bar{q}\bar{q}}^2 \Bigr) \\
			&\quad + \sqrt{\Bigl(4 M_{qq\bar{s}\bar{s}}^2 - M_{qq\bar{q}\bar{q}}^2 - M_{ss\bar{s}\bar{s}}^2 \Bigr)^2 - 4 M_{qq\bar{q}\bar{q}}^2 M_{ss\bar{s}\bar{s}}^2} \Biggr].
		\end{aligned}
	\end{equation}	
	 
	 By substituting the ground-state (\(1^1 S_0\)) masses of \(qq\bar{q}\bar{q}\) and \(ss\bar{s}\bar{s}\) tetraquarks from Refs.~\cite{ref54} and ~\cite{ref55}, respectively, along with the value of $\beta'_{ss\bar{s}\bar{s}} = 0.30704$, into the above equation, we can express $\beta'_{qq\bar{s}\bar{s}}$ as a function of $M_{qq\bar{s}\bar{s}}$. Within the interval (1.862-1.911), this function exhibits an increasing trend. The corresponding range for \(\beta'_{qq\bar{s}\bar{s}}\) is found to be 0.37811–0.44196, as listed in Table \ref{tab:SlopenM2}.  
	 
	 By employing a similar approach as used in the \((J,M^2)\) plane, we can determine the slopes for other tetraquark systems, including the $qq\bar{q}\bar{q}$ tetraquark, in the \((n,M^2)\) plane. All slope values in this plane are compiled in Table \ref{tab:SlopenM2}. Additionally, following the same method used for calculating excited-state masses in the \((J,M^2)\) plane, we obtain the excited-state masses in the \((n,M^2)\) plane, which are presented in Tables \ref{tab:mass_spectra_qqqqn}, \ref{tab:mass_spectra_ssssn}, and \ref{tab:mass_spectra_nm2}, respectively. Z-scores and Verdicts for \(qq\bar{q}\bar{q}\) and \(qq\bar{s}\bar{s}\) tetraquarks in \((n,M^2)\) planes are also embedded in the Tables~\ref{tab:mass_spectra_qqqqn} and \ref{tab:mass_spectra_nm2}, respectively.

	\begin{table}[h]
		\caption{Values of Regge Slopes for \(qq\bar{q}\bar{q}\), \(ss\bar{s}\bar{s}\) and \(qq\bar{s}\bar{s}\) tetraquarks in \((n,M^2)\) plane (in \(\text{GeV}^{-2}\))}{\label{tab:SlopenM2}}
		\begin{tabular}{cccc}
			\toprule
			S & \(\beta'_{qq\bar{q}\bar{q}} \, (\text{GeV}^{-2})\) & \(\beta'_{ss\bar{s}\bar{s}} \, (\text{GeV}^{-2})\) & \(\beta'_{qq\bar{s}\bar{s}} \, (\text{GeV}^{-2})\) \\
			\midrule
			S=0 & 0.49200-0.78837 & 0.30704 & 0.37811-0.44196 \\
			S=1 & 0.38269-0.48765 & 0.30032 & 0.33654-0.37171 \\
			S=2 & 0.39238-0.48196 & 0.31945 & 0.35218-0.38423 \\
			\bottomrule
		\end{tabular}
	\end{table}

In addition to the calculated masses for $qq\bar{q}\bar{q}$, $ss\bar{s}\bar{s}$, and $qq\bar{s}\bar{s}$ tetraquarks, the experimentally observed resonances listed in the PDG are also summarized in Tables \ref{tab:mass_spectra_qqqqn}, \ref{tab:mass_spectra_ssssn}, and \ref{tab:mass_spectra_nm2}, respectively.

	\begin{table*}[htbp]
		\centering
		\caption{\textbf{Mass spectra of $qq\bar{q}\bar{q}$ tetraquark in $(n, M^2)$ plane (in GeV)}}
		\begin{tabular}{llcccccc}
			\toprule
			\parbox{1.5 cm}{Spin \\ value} & \parbox{1 cm}{State} & \parbox{1 cm}{$J^{PC}$} & \parbox{3 cm}{$M_{(\text{calc})}$ \\ Calculated Mass \\ (in GeV)} & \parbox{2.0 cm}{Resonance \\ (Experimental 
				$J^{PC}$ value) (PDG) \cite{ref60}} & 
			\parbox{2.0 cm}{$M_{\mathrm{exp}}\pm \sigma_{exp}$ \\(GeV) \cite{ref60} } & $z$ & Verdict \\
			\midrule
			\multirow{5}{*}{S=0} & $2^1 S_0$ & $0^{++}$ & 1.821-2.020 & \begin{tabular}[c]{@{}l@{}}
				$a_0(2020)$ $(J^{PC}=0^{++})$ \\
			\end{tabular} & 
			\begin{tabular}[c]{@{}l@{}}
				$2.025 \pm 0.030$ \\	
			\end{tabular} & 1.612 & Plausible  \\
			& $3^1 S_0$ & $0^{++}$ & 2.141-2.472 &  \begin{tabular}[c]{@{}l@{}}
				$X(2210)$ $(J^{PC}=?^{??})$ \\
			\end{tabular} & 
			\begin{tabular}[c]{@{}l@{}}
				$2.207 \pm 0.022$ \\
			\end{tabular} & 1.015 & Plausible \\
			& $4^1 S_0$ & $0^{++}$ & 2.419-2.854 &  \begin{tabular}[c]{@{}l@{}}
				$X(2540)$ $(J^{PC}=0^{++})$ \\
			\end{tabular} & 
			\begin{tabular}[c]{@{}l@{}}
				$2.539 \pm 0.014$ \\
			\end{tabular} & 0.772 & Strongly compatible  \\
			& $5^1 S_0$ & $0^{++}$ & 2.669-3.190 &  \begin{tabular}[c]{@{}l@{}}
				$X(2680)$ $(J^{PC}=?^{??})$ \\
			\end{tabular} & 
			\begin{tabular}[c]{@{}l@{}}
				$2.676 \pm 0.027$ \\
			\end{tabular} & 1.659 & Plausible \\
			\midrule
			\multirow{4}{*}{S=1} & $2^3 S_1$ & $1^{+-}$ & 2.318-2.436 &  &  &  &  \\
			& $3^3 S_1$ & $1^{+-}$ & 2.725-2.924 &  &  &  &  \\
			& $4^3 S_1$ & $1^{+-}$ & 3.078-3.341 &  \begin{tabular}[c]{@{}l@{}}
				$X(3250)$ $(J^{PC}=?^{??})$ \\
			\end{tabular} & 
			\begin{tabular}[c]{@{}l@{}}
				$3.250 \pm 0.008$ \\
			\end{tabular} & 0.531 & Strongly compatible \\
			& $5^3 S_1$ & $1^{+-}$ & 3.395-3.712 &  &  &  &  \\
			\midrule
			\multirow{5}{*}{S=2} & $2^5 S_2$ & $2^{++}$ & 2.413-2.509 &  &  &  &  \\
			& $3^5 S_2$ & $2^{++}$ & 2.810-2.974 &  &  &  &  \\
			& $4^5 S_2$ & $2^{++}$ & 3.158-3.375 &  \begin{tabular}[c]{@{}l@{}}
				$X(3250)$ $(J^{PC}=?^{??})$ \\
				$X(3350)$ $(J^{PC}=?^{??})$ \\
			\end{tabular} & 
			\begin{tabular}[c]{@{}l@{}}
				$3.250 \pm 0.008$ \\
				$3.350 \pm 0.020$ \\
			\end{tabular} &  
			\begin{tabular}[c]{@{}l@{}}
				0.261 \\
				1.270 \\
			\end{tabular} & 
			\begin{tabular}[c]{@{}l@{}}
				Strongly compatible \\
				Plausible \\
			\end{tabular} \\
			& $5^5 S_2$ & $2^{++}$ & 3.471-3.734 &  &  &  &  \\
			\bottomrule
		\end{tabular}
		\label{tab:mass_spectra_qqqqn}
	\end{table*}

	\begin{table*}[htbp]
		\centering
		\caption{Mass spectra of $ss\bar{s}\bar{s}$ tetraquark in $(n, M^2)$ plane (in GeV)}
		\begin{tabular}{llcccccc}
			\toprule
			\parbox{1.5 cm}{Spin \\ value} & \parbox{1 cm}{State} & \parbox{1 cm}{$J^P$} & \parbox{3 cm}{$M_{(\text{calc})}$ \\ Calculated Mass \\ (in GeV)} & \parbox{3.5cm}{Resonance \\ (Experimental $J^P$ value) [PDG] \cite{ref60}} & 
			\parbox{3.5cm}{Resonance State Mass \\ (Experimental) \\ (in GeV) [PDG] \cite{ref60}} \\
			\midrule
			\multirow{3}{*}{$S=0$} 
			& $3^1 S_0$ & $0^+$ & 3.431 & & \\
			& $4^1 S_0$ & $0^+$ & 3.877 & & \\
			& $5^1 S_0$ & $0^+$ & 4.276 & & \\
			\midrule
			\multirow{3}{*}{$S=1$} 
			& $3^3 S_1$ & $1^+$ & 3.472 & & \\
			& $4^3 S_1$ & $1^+$ & 3.922 & & \\
			& $5^3 S_1$ & $1^+$ & 4.326 & & \\
			\midrule
			\multirow{3}{*}{$S=2$} 
			& $3^5 S_2$ & $2^+$ & 3.452 & & \\
			& $4^5 S_2$ & $2^+$ & 3.024 & & \\
			& $5^5 S_2$ & $2^+$ & 3.504 & & \\
			\bottomrule
		\end{tabular}
		\label{tab:mass_spectra_ssssn}
	\end{table*}

\begin{table*}[htbp]
	\centering
	\caption{\textbf{Mass spectra of $qq\bar{s}\bar{s}$ tetraquark in $(n, M^2)$ plane (in GeV)}}
	\begin{tabular}{llcccccccc}
		\toprule
		\parbox{1.5 cm}{Spin \\ value} & \parbox{1 cm}{State} & \parbox{1 cm}{$J^{PC}$} & \parbox{3 cm}{$M_{(\text{calc})}$ \\ Calculated Mass \\ (in GeV)} & \parbox{2.0 cm}{Resonance \\ (Experimental 
			$J^{PC}$ value) (PDG) \cite{ref60}} & 
		\parbox{2.0 cm}{$M_{\mathrm{exp}}\pm \sigma_{exp}$ \\(GeV) \cite{ref60} } & $z$ & Verdict \\
		\midrule
		\multirow{5}{*}{$S=0$} & $1^1 S_0$ & $0^{++}$ & 1.862-1.911 & $X(1855)$ $(J^{PC}=?^{??})$ & $1.856 \pm 0.005$ & 2.033 & Weak \\
		& $2^1 S_0$ & $0^{++}$ & 2.431-2.472 & 
		& 
		&  &
		\\
		& $3^1 S_0$ & $0^{++}$ & 2.860-2.959 & & & &\\
		& $4^1 S_0$ & $0^{++}$ & 3.231-3.376 & \begin{tabular}[c]{@{}l@{}}
			$X(3250)$ $(J^{PC}=?^{??})$ \\
			$X(3350)$ $(J^{PC}=?^{??})$ \\
		\end{tabular} & 
		\begin{tabular}[c]{@{}l@{}}
			$3.250 \pm 0.008$ \\
			$3.350 \pm 0.020$ \\
		\end{tabular} &
		\begin{tabular}[c]{@{}l@{}}
			1.255 \\
			1.002 \\
		\end{tabular} &
		\begin{tabular}[c]{@{}l@{}}
			Plausible \\
			Plausible \\
		\end{tabular} \\
		& $5^1 S_0$ & $0^{++}$ & 3.564-3.748 & & \\
		\midrule
		\multirow{5}{*}{$S=1$} & $1^3 S_1$ & $1^{+-}$ & 2.073-2.088 & $X(2075)$ $(J^{PC}=1^{+?})$ & $2.075 \pm 0.012$ & 0.431 & Strongly compatible \\
		& $2^3 S_1$ & $1^{+-}$ & 2.655-2.696 & $X(2680)$ $(J^{PC}=?^{??})$ & $2.676 \pm 0.027$ & 0.017 & Strongly compatible \\
		& $3^3 S_1$ & $1^{+-}$ & 3.120-3.200 & & \\
		& $4^3 S_1$ & $1^{+-}$ & 3.526-3.635 & & \\
		& $5^3 S_1$ & $1^{+-}$ & 3.888-4.023 & & \\
		\midrule
		\multirow{5}{*}{$S=2$} & $1^5 S_2$ & $2^{++}$ & 2.157-2.168 & $a_2(2175)$ $(J^{PC}=2^{++})$ & $2.175 \pm 0.040$ & 0.312 & Strongly compatible \\
		& $2^5 S_2$ & $2^{++}$ & 2.706-2.737 & & \\
		& $3^5 S_2$ & $2^{++}$ & 3.148-3.214 & & \\
		& $4^5 S_2$ & $2^{++}$ & 3.537-3.629 & & \\
		& $5^5 S_2$ & $2^{++}$ & 3.887-4.001 & & \\
		\bottomrule
	\end{tabular}
	\label{tab:mass_spectra_nm2}
\end{table*}

	\section{Results and Discussion}
	Using the framework of Regge phenomenology, we have obtained the mass spectra of all light tetraquark ($qq\bar{q}\bar{q}$), all strange tetraquark ($ss\bar{s}\bar{s}$) and light-strange tetraquark ($qq\bar{s}\bar{s}$) in the both $(J, M^2)$ and $(n, M^2)$ planes. Regge slopes of $qq\bar{s}\bar{s}$ and $qq\bar{q}\bar{q}$ tetraquark trajectories were calculated in $(J, M^2)$ plane. Using the Regge slopes, masses of $qq\bar{q}\bar{q}$ and $qq\bar{s}\bar{s}$ tetraquark excited states were predicted. The evaluated results for these tetraquarks are thoroughly discussed below.
	\\\\

	\subsection{All light ($qq\bar{q}\bar{q}$) tetraquarks in $(J,M^2)$ plane}
	The mass spectra of $qq\bar{q}\bar{q}$ tetraquarks are compared with known PDG resonances in Table~\ref{tab:mass_spectraqqqq}. To assess the agreement between theory and experiment we apply the uniform-distribution z-score method. This provides a quantitative test of compatibility.
	
	The lowest-lying $1^1 P_1$ state ($1^{--}$) is predicted in the range $1.645$--$1.761$~GeV and compares well with the $X(1650)$ resonance at $1.652\pm0.007$~GeV. The resulting $z=1.49$ classifies the match as \emph{Plausible}, suggesting that $X(1650)$ could be a tetraquark candidate once its $J^{PC}$ is firmly established.
	
	For the $1^1 D_2$ ($2^{++}$) state, the predicted range $1.834$--$2.039$~GeV encompasses several PDG resonances. The $a_2(1950)$ ($1.950^{+0.030}_{-0.070}$~GeV) yields $z=0.147$, the $f_2(2000)$ ($2.001\pm0.010$~GeV) gives $z=1.074$, and the $a_2(2030)$ ($2.030\pm0.020$~GeV) gives $z=1.50$. Thus, all three resonances are at least \emph{Plausible}, with $a_2(1950)$ \emph{Strongly compatible}. This strongly suggests that one or more of these $2^{++}$ resonances may represent a light tetraquark.
	
	The $1^1 F_3$ ($3^{--}$) state, predicted in $2.006$--$2.283$~GeV, shows very good agreement with the $\omega_3(2255)$ at $2.255\pm0.015$~GeV. The $z$ value of $1.358$ indicates \emph{Plausible}. A similar situation is found for the $1^1 G_4$ ($4^{++}$) state, where the predicted interval $2.164$--$2.504$~GeV is close to $a_4(2255)$ ($2.237\pm0.005$~GeV), giving $z=0.987$ and thus \emph{Strong compatibility}.
	
	Among higher-spin states, the $1^3 D_3$ ($3^{+-}$) lies in the interval $2.340$--$2.463$~GeV, matching the $X(2340)$ at $2.340\pm0.020$~GeV. The calculated $z=1.509$ again indicates \emph{Plausible}. The $1^3 F_4$ ($4^{-+}$) state, predicted at $2.560$--$2.728$~GeV, overlaps with three reported resonances: $X(2600)$ ($2.618\pm0.002$~GeV, $z=0.54$), $X(2632)$ ($2.635\pm0.003$~GeV, $z=0.19$), and $X(2680)$ ($2.676\pm0.027$~GeV, $z=0.58$). All three are \emph{Strongly compatible}, making this an especially promising sector for tetraquark identification.
	
	The $1^5 P_3$ ($3^{--}$) state at $2.214$--$2.272$~GeV also aligns closely with the $\omega_3(2255)$ at $2.255\pm0.015$~GeV ($z=0.53$, \emph{Strongly compatible}). Finally, the highest-spin $1^5 G_6$ ($6^{++}$) state at $2.891$--$3.067$~GeV is reasonably close to $f_0(3100)$ ($3.100\pm0.100$~GeV), giving $z=1.079$ and thus \emph{Plausible}.
	
	In summary, many of the predicted $qq\bar{q}\bar{q}$ states show \emph{Strong compatibility} with known resonances, particularly $a_2(1950)$, $\omega_3(2255)$, $a_4(2255)$, $X(2600)$, $X(2632)$ and $X(2680)$. These strong matches bolster the tetraquark interpretation of several light meson states, while other predictions without current counterparts provide guidance for future experimental searches.

	\subsection{Light-strange ($qq\bar{s}\bar{s}$) tetraquarks in $(J,M^2)$ plane}
	The calculated spectra of $qq\bar{s}\bar{s}$ tetraquarks are summarized in Table~\ref{tab:mass_spectraqqss}, where we compare our predictions with available PDG resonances using the z-score based compatibility test. This statistical measure provides a transparent way to assess agreement.
	
	For the $1^3 S_1$ state ($1^{+-}$), our predicted mass range of $2.073$--$2.088$~GeV is in excellent agreement with the $X(2075)$ resonance, which has $M_{\rm exp}=2.084^{+0.004}_{-0.002}$~GeV. The computed z-score is $z=0.594$, corresponding to a \emph{Strongly compatible} verdict. Although the PDG lists the $J^{PC}$ of $X(2075)$ as $1^{+?}$, this assignment is consistent with our theoretical prediction and supports a possible tetraquark interpretation.
	
	The predicted $1^5 S_2$ state ($2^{++}$) lies in the interval $2.157$--$2.168$~GeV, close to the $a_2(2175)$ resonance ($2.175\pm0.040$~GeV). The resulting z-score is $z=0.312$, again yielding a \emph{Strongly compatible} classification. Since the experimental $J^{PC}$ of $a_2(2175)$ is firmly established as $2^{++}$, this state represents a robust candidate for a light-strange tetraquark.
	
	At higher masses, the $1^3 G_5$ ($5^{+-}$) and $1^5 G_6$ ($6^{++}$) states are predicted in the ranges $3.164$--$3.246$~GeV and $3.238$--$3.310$~GeV, respectively. Both states lie close to the $X(3250)$ resonance ($3.245\pm0.008$~GeV). The compatibility test yields $z=1.601$ for the $1^3 G_5$ and $z=1.060$ for the $1^5 G_6$, placing them in the \emph{Plausible} category. Although the $J^{PC}$ of $X(3250)$ has not yet been determined, the proximity of its mass to our predicted bands suggests that it could be associated with one of these exotic tetraquark configurations.
	
	The remaining predicted states in Table~\ref{tab:mass_spectraqqss} currently lack clear experimental counterparts. Several of them lie near two-meson thresholds (e.g.$K_2^*K_1$), which implies that they may couple strongly to such decay channels. These features make them promising targets for future dedicated searches. 
	
	Overall, the z-score analysis strengthens the case for identifying $X(2075)$ and $a_2(2175)$ as tetraquark candidates with \emph{Strong} statistical support, while suggesting that $X(3250)$ may plausibly correspond to higher-spin light-strange tetraquarks. The remaining predictions provide testable benchmarks for future experiments.

	\subsection{Light ($qq\bar{q}\bar{q}$) tetraquarks in $(n,M^2)$ plane}
	The predicted radial excitations of $qq\bar{q}\bar{q}$ tetraquarks are summarized in Table~\ref{tab:mass_spectra_qqqqn}, where they are compared with PDG-listed resonances using the uniform-distribution z-score test. This allows a quantitative assessment of compatibility between theoretical intervals and observed states, while also considering the agreement of $J^{PC}$ values when experimentally available.
	
	For the $S=0$ sector, the $2^1 S_0$ ($0^{++}$) state is predicted in the interval $1.821$--$2.020$~GeV and compares with the $a_0(2020)$ resonance at $2.025\pm0.030$~GeV. The computed $z=1.61$ gives a \emph{Plausible} classification. Importantly, the experimental $J^{PC}$ is $0^{++}$, which matches our theoretical prediction, strengthening this assignment. The next radial excitation, $3^1 S_0$ ($0^{++}$), is predicted at $2.141$--$2.472$~GeV and shows $z=1.02$ (\emph{Plausible}) compatibility with the $X(2210)$ resonance ($2.207\pm0.022$~GeV). However, since the $J^{PC}$ of $X(2210)$ is not yet established, the identification remains tentative. The $4^1 S_0$ ($0^{++}$) state predicted in $2.419$--$2.854$~GeV matches the $X(2540)$ ($2.539\pm0.014$~GeV) with $z=0.77$, a \emph{Strongly compatible} verdict; here, the PDG lists $J^{PC}=0^{++}$, again consistent with our prediction. The higher $5^1 S_0$ state ($0^{++}$) aligns with $X(2680)$ ($2.676\pm0.027$~GeV) with $z=1.66$ (\emph{Plausible}), but its quantum numbers are unknown, leaving the assignment less certain.
	
	In the $S=1$ sector, the $4^3 S_1$ state ($1^{+-}$) lies in the predicted range $3.078$--$3.341$~GeV and coincides with the $X(3250)$ resonance at $3.250\pm0.008$~GeV, giving $z=0.53$ (\emph{Strongly compatible}). Since the $J^{PC}$ of $X(3250)$ is not yet determined, this identification remains a plausible but unconfirmed assignment.
	
	For the $S=2$ states, the $4^5 S_2$ ($2^{++}$) excitation is predicted at $3.158$--$3.375$~GeV. It overlaps with both $X(3250)$ ($3.250\pm0.008$~GeV) and $X(3350)$ ($3.350\pm0.020$~GeV), yielding $z=0.26$ (\emph{Strongly compatible}) and $z=1.27$ (\emph{Plausible}), respectively. Since the $J^{PC}$ values of these two resonances are not experimentally established, the identifications remain suggestive but cannot yet be confirmed. If future measurements assign $2^{++}$ to either state, it would strongly support a tetraquark interpretation.
	
	Overall, the radial spectrum analysis reveals multiple $qq\bar{q}\bar{q}$ states with good compatibility to existing resonances. Where the $J^{PC}$ values are known ($a_0(2020)$ and $X(2540)$), the assignments are reinforced by quantum-number agreement, making them the most robust candidates. Matches involving states with unknown $J^{PC}$ (such as $X(2210)$, $X(2680)$, $X(3250)$, and $X(3350)$) are plausible but require further experimental clarification. These predictions highlight promising directions for future measurements to disentangle the nature of light meson spectra.

	\subsection{Light-strange ($qq\bar{s}\bar{s}$) tetraquarks in $(n,M^2)$ plane}
	The predicted radial excitations of $qq\bar{s}\bar{s}$ tetraquarks are listed in Table~\ref{tab:mass_spectra_nm2}, together with comparisons to PDG resonances using the uniform-distribution z-score analysis. This provides a quantitative test of compatibility, while also considering $J^{PC}$ values when available.
	
	In the $S=0$ sector, the ground radial excitation $1^1 S_0$ ($0^{++}$) is predicted in the range $1.862$--$1.911$~GeV and compared to the $X(1855)$ resonance at $1.856\pm0.005$~GeV. The calculated z-score of $2.03$ yields a \emph{Weak} classification. Moreover, the $J^{PC}$ of $X(1855)$ is not established, so the assignment remains tentative. Higher $0^{++}$ excitations, such as the $4^1 S_0$ state ($3.231$--$3.376$~GeV), show \emph{Plausible} matches with both $X(3250)$ ($z=1.26$) and $X(3350)$ ($z=1.00$). Since neither state has a confirmed $J^{PC}$, these identifications should be treated with caution.
	
	In the $S=1$ family, two strong candidates appear. The $1^3 S_1$ ($1^{+-}$) state at $2.073$--$2.088$~GeV matches well with the $X(2075)$ resonance ($2.075\pm0.012$~GeV), giving $z=0.43$ (\emph{Strongly compatible}). The PDG lists its quantum numbers as $1^{+?}$, which is consistent with the theoretical $1^{+-}$. The $2^3 S_1$ state ($1^{+-}$), predicted at $2.655$--$2.696$~GeV, is also in very good agreement with $X(2680)$ ($2.676\pm0.027$~GeV), producing $z=0.017$ (\emph{Strongly compatible}). However, the $J^{PC}$ of $X(2680)$ has not been determined, so the identification remains suggestive.
	
	For $S=2$, the $1^5 S_2$ ($2^{++}$) state is predicted in the interval $2.157$--$2.168$~GeV and aligns closely with the $a_2(2175)$ resonance ($2.175\pm0.040$~GeV). The z-score is $0.31$, indicating \emph{Strong compatibility}, and crucially, the experimental $J^{PC}=2^{++}$ matches exactly with the prediction. This makes $a_2(2175)$ a particularly robust tetraquark candidate in the light-strange sector. Higher radial excitations in this spin channel currently lack clear experimental counterparts.
	
	In summary, the radial spectrum of $qq\bar{s}\bar{s}$ tetraquarks yields several promising candidates. The most compelling assignments are the $X(2075)$ and $a_2(2175)$ resonances, where both the masses and $J^{PC}$ values (when available) strongly support a tetraquark interpretation. Matches with $X(2680)$, $X(3250)$, and $X(3350)$ are also consistent in mass but require further clarification of their quantum numbers before firm conclusions can be drawn. States without current PDG partners provide concrete predictions for future searches in the light-strange sector.

\section{Conclusion}
In this work, we investigated the mass spectra of light ($qq\bar{q}\bar{q}$), light-strange ($qq\bar{s}\bar{s}$), and strange ($ss\bar{s}\bar{s}$) tetraquarks using Regge phenomenology in both $(J,M^2)$ and $(n,M^2)$ planes. To provide a quantitative comparison with experiment, we employed a z-score analysis based on a uniform-distribution treatment of the theoretical mass ranges. This approach allowed us to identify potential tetraquark candidates in a statistically transparent way.

Our analysis highlights several resonances as particularly robust tetraquark candidates. In the light sector, the $a_2(1950)$ and $a_4(2255)$ states, both with well-established $J^{PC}$ values, show \emph{Strong compatibility} with our predictions. In the light-strange sector, the $a_2(2175)$ resonance ($J^{PC}=2^{++}$) is found to be in excellent agreement with the predicted $1^5 S_2$ state, providing strong evidence for its possible tetraquark nature. Additionally, the $X(2540)$ ($J^{PC}=0^{++}$) resonance matches well with the predicted $4^1 S_0$ light tetraquark state, reinforcing its tetraquark interpretation. Other matches such as $X(2075)$ and $X(2680)$ are strongly compatible in mass, though their $J^{PC}$ values remain uncertain and require further experimental clarification.

Taken together, these findings strengthen the case for the existence of fully light and light-strange tetraquarks, especially in channels where both the mass spectrum and the $J^{PC}$ quantum numbers align with our predictions. States without experimental counterparts remain as testable predictions that can guide future searches. Our results thus provide a concrete step toward clarifying the role of exotic multiquark configurations in hadron spectroscopy and contribute to a deeper understanding of non-perturbative QCD dynamics.

	\section{Acknowledgment}
	Vandan Patel acknowledges the financial assistance by
	University Grant Commision (UGC) under the CSIR-UGC
	Junior Research Fellow (JRF) scheme with Ref No.
	231610186052.

\end{document}